\documentclass[letter]{aa}     
\usepackage{graphicx}
\usepackage{txfonts}
\usepackage{lipsum}
\usepackage{subcaption}                                  
\usepackage{lscape}                                      
\usepackage{placeins}          
\usepackage[breaklinks=true]{hyperref} 

\begin{document}

\title{Pre-perihelion detection of a wobbling \textbf{high-latitude} jet in the interstellar comet 3I/ATLAS \thanks{Based on observations made with the Two-meter Twin Telescope (TTT)}}

   \author{M. Serra-Ricart\inst{1,2,3}
        \and J. Licandro\inst{2,3}
        \and M. R. Alarcon\inst{1,2,3}  
         }
   \authorrunning{M. Serra-Ricart et al.}
   \titlerunning{Pre-perihelion jet detection in 3I/ATLAS} 
   \offprints{M. Serra-Ricart, \email{miquel@lightbridges.es}}
   \institute{
        Light Bridges, Observatorio Astronómico del Teide. 
        Carretera del Observatorio del Teide, s/n, Güímar, Tenerife, Spain
        \and Instituto de Astrofísica de Canarias (IAC), C/Vía Láctea s/n, 38205 La Laguna, Tenerife, Spain
        \and Departamento de Astrofísica, Universidad de La Laguna, Avda. Astrofísico Francisco Sánchez, 38206 La Laguna, Tenerife, Spain\\}

\date{Received 11 November 2025 / Accepted 04 December 2025}

\abstract
{}
{We investigate the pre-perihelion rotational parameters of the nucleus 
of 3I/ATLAS modeling jet structures observed in the inner coma.
} 
{The comet was extensively monitored on 37 nights between 2025, July 2 and September 5, using the imaging capabilities of the Two-meter Twin Telescope (TTT) at the Teide Observatory (Tenerife, Canary Islands, Spain). To enhance the visibility of potential fine-scale structures in the inner coma of 3I/ATLAS, a Laplacian-filtering technique was applied to the reduced and combined images.}
{We present observations of the detection of a faint high-latitude jet in the inner coma of comet 3I/ATLAS that coincides with the broad plume detected in visible images along PA $280\pm10\degr$. 
A detailed analysis shows that the jet was clearly detected on seven nights (2025, August~3, 5, 18, 19, 21, 24, and~29). The jet maintains an almost, though not perfectly, constant position angle (PA) throughout these epochs. 
High-precision PA measurements at a projected distance of 6000~km from the cometary optocenter reveal a periodic modulation centered at $\sim280\degr$, consistent with a  high-latitude jet undergoing precessional motion around the sky-projected spin axis of the nucleus. This is the first periodic jet-angle modulation detected in an interstellar comet.
The derived periodicity of $7.74 \pm 0.35$~h may imply a nucleus rotation period of $P_{\mathrm{rot}} = 15.48 \pm 0.70$~h if the jet originates from a single active source near one of the poles. This value is slightly shorter than the period of $P_{\mathrm{rot}} = 16.79 \pm 0.23$~h derived from the photometric time series.
From the measured PA range, the sky-projected orientation of the spin axis is derived as $\mathrm{PA} = 280.7 \pm 0.2^{\circ}$.
}
{}


\maketitle
\nolinenumbers 

\section{Introduction}
\label{sec:intro}
Cometary jets -- localized, collimated outflows of gas and dust -- are key tracers of nucleus activity and surface inhomogeneity \citep{Sekanina1986, Sekanina2004, Samarasinha2014}. Their detection provides constraints on the spin-axis orientation, rotation state, and nongravitational forces that drive cometary evolution \citep{Schleicher2004, Kramer2019}. Image-enhancement techniques such as Laplace \citep{SerraRicart2015Rotation} or Larson–Sekanina filtering \citep{LarsonSekanina1984_HalleyI} and azimuthal average subtraction are commonly used to isolate faint, anisotropic structures from the bright coma background \citep{Boehnhardt1994}. Measuring the position angle (PA) of individual jets and tracking their evolution enables modeling of the nucleus geometry and identification of discrete active regions.
The recently discovered interstellar comet 3I/ATLAS (C/2025 N1 hereafter 3I, \citealt{Tonry2025ATLAS}) presents a unique opportunity to extend jet studies beyond the Solar System. 

As only the third confirmed interstellar object -- after 1I/‘Oumuamua and 2I/Borisov -- 3I likely retains pristine volatile and dust from the origin of a exoplanetary system that has likely undergone long-term irradiation by galactic cosmic rays in the interstellar space and some additional irradiation in other stellar systems. Such processing can significantly modify the physical and chemical properties of surface materials and should be distinguished from solar-driven alteration processes \citep{2025arXiv251026308M,2025ApJ...993L..31Y}.
Detecting a jet in 3I would provide the first evidence of localized outgassing in an interstellar nucleus, allowing for a direct comparison with activity mechanisms observed in Solar System comets. Such a detection would constrain the orientation and rotation state of the nucleus, quantify possible nongravitational accelerations caused by anisotropic mass loss, and reveal the onset of activity as a function of heliocentric distance, thereby identifying the dominant volatile species (e.g., CO, CO$_2$, and H$_2$O) responsible for outflow initiation. Characterizing jets in 3I thus represents a rare opportunity to investigate the physical behavior of a pristine body formed in another planetary system.

\section{Observations} 
\label{sec:observations}
3I was extensively monitored on 37 nights between 2025, July 2 and September 5 (see table \ref{table:obs}), 
using the imaging capabilities of the Two-meter Twin Telescope (TTT\footnote{\url{https://ttt.iac.es/}}). TTT is located at the Teide Observatory on the island of Tenerife (Canary Islands, Spain). Currently, it includes two 0.80-m telescopes (TTT1 and TTT2) and a 2.0-m telescope (TTT3). All images presented in the paper were observed with TTT3. 
TTT3 is a 2.0-m $f$/6 Ritchey-Chrétien telescope that is currently in its commissioning phase. 3I images were obtained with COLORS, a 2k$\times$2k camera mounted at the Nasmyth 2 focus, equipped with a back-illuminated 13.5~$\mu$m pixel$^{-1}$ BEX2-DD CCD sensor, resulting in a field of view of 7.85$'$$\times$7.85$'$ and a plate scale of 0.23$''$~pixel$^{-1}$. Jet images were taken using the Sloan-$g'$ filter. All the images were bias, dark, and flat-field corrected in the standard way. A  WCS (world coordinate system) solution was obtained using Astrometry.net \citep{2010AJ....139.1782L} against Gaia DR2 and all comet images are aligned from the comet’s optocenter.

\begin{figure*}[h!]
    \centering
     \resizebox{18cm}{12.67cm}
    {\includegraphics {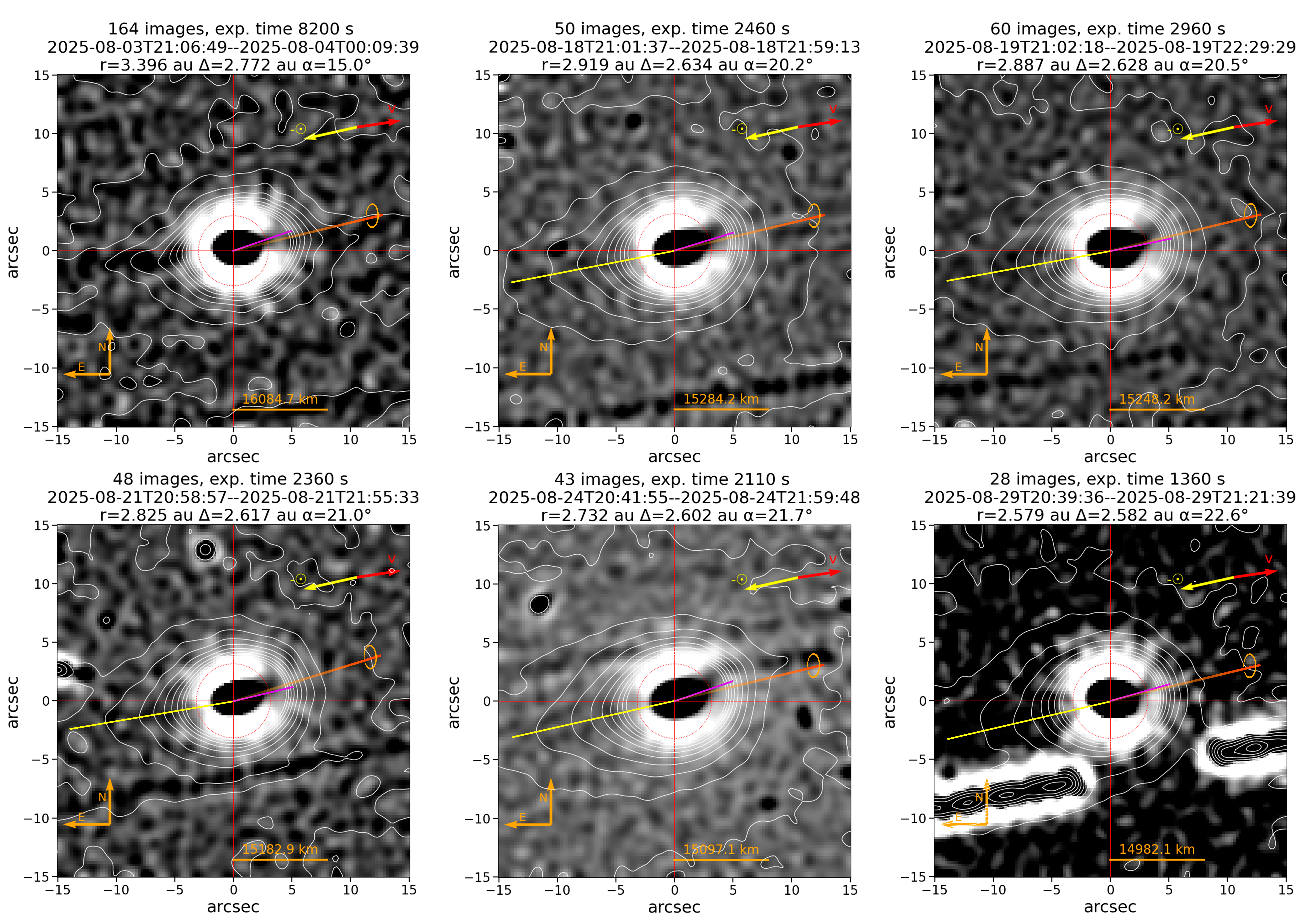}}
    \caption{
    Laplacian-filtered images of the inner coma of the interstellar comet 3I/ATLAS (C/2025~N1). 
    Thin purple lines indicate the PA of the detected jet (median value when multiple measurements are available), calculated at a projected distance of 6000~km from the comet’s optocenter (dotted red circle). For each frame, the observation date (YYYY–MM–DD) and the start and end times in UTC are shown above the panels, together with the total number of sidereal-tracking exposures and the cumulative integration time. 
    The projected velocity vector (red arrow) and the antisolar direction (yellow arrow) are marked, as well as the image scale and orientation. 
   Degraded orange lines denote the projected spin-axis direction (PA~=~$280.7\pm0.2^{\circ}$), derived from mean value of the measured jet PAs, while yellow lines trace the tail direction. Red crosshairs mark the comet’s optocenter. Isophotal contours of the original (unfiltered) images are overplotted using ten logarithmically spaced levels between the 20th and 95th percentiles of pixel intensity in each frame. The jet refers to the narrow, linear, black feature extending roughly northwest in the filtered images. It originates on the sunward side of the nucleus, in contrast to the dust tail, which points in the antisolar direction.}

      \label{fig:jets}
\end{figure*}

\section{Methods}
\label{sec:methods}

To enhance the visibility of potential fine-scale structures in the inner coma of 3I, a Laplacian-filtering technique was applied to the reduced and combined images. The Laplacian operator highlights spatial curvature, and therefore efficiently suppresses the smooth, steep radial gradient of the coma, while enhancing localized features such as narrow jets or filaments. Unlike azimuthal-average subtraction or the Sekanina–Larson residual technique, the Laplacian filter does not rely on any assumed symmetry of the coma, nor does it require iterative rotational subtraction. This makes the method more robust in the presence of multiple or curved jets and less sensitive to small cometary optocenter errors. The resulting Laplacian-filtered frames provide clean, high-contrast representations of any detected jet morphology, from which accurate position-angle profiles can be extracted for quantitative analysis following the procedure described by \citet{SerraRicart2015Rotation}.
The detection and analysis of jet features were often hampered by low signal-to-noise ratios, imperfect tracking, confusion with background stars, or poor seeing conditions.
After careful inspection and processing of all datasets (see Table~\ref{table:obs}), only seven observing nights (August 3, 5, 18, 19, 21, 24, and 29) revealed evidence of a faint yet clearly defined jet structure. 
The corresponding Laplace-filtered frames are displayed in Fig.~\ref{fig:jets} (for clarity the August 5 frame is not included). In each panel, the PA of the jet -- measured at a projected optocenter distance of 6000~km -- is indicated by purple lines (median value when multiple measurements are available). Isophotal contours from the original, unfiltered images are superimposed for reference. 

In these Laplace-enhanced images, black regions correspond to sharp structural gradients such as jets, tail streamers, shells, or point-like sources, whereas white regions represent flat areas of the coma. The “jet” refers to the narrow, linear, black feature extending roughly northwest in the filtered images (see figures \ref{fig:jets} and \ref{fig:filunfil}). It should be noted that the Laplacian filter does not preserve absolute flux; 
only morphological information and the spatial distribution of features can be reliably inferred from the processed data.

To determine the jet PAs, we applied the methodology described by \citet{SerraRicart2015Rotation}.  
Each original image was first processed with a Laplacian filter, and then transformed into polar coordinates (PA versus radius, see figure \ref{fig:fitting}).  
If the jet represents the two-dimensional projection of an Archimedean spiral, the PA-radius relation becomes approximately linear near the nucleus, transforming the jet into a straight feature in the polar frame.  
From the filtered polar images, PA profiles were extracted at different radial distances (within $\pm250$~km around 6000~km from the optocenter).  
For each radius, the jet PA was measured by fitting a parabola to the local intensity minimum within the jet region.  
A linear regression of the resulting PA measurements  yielded the final PA of the jet (at $r = 6000$~km) and its uncertainty for each epoch (see Fig. \ref{fig:fitting}).  
This procedure was applied to all nights in which the jet was clearly detected (see Fig.~\ref{fig:jets}).  
To improve temporal sampling, several random subsets of 1200~s exposures were independently analyzed for each night when sufficient data were available. 

\section{Results and discussion}
\label{sec:results}

\subsection{Coma of 3I}
\label{sec:results.coma}

As noted by several observers \citep{Bolin2025, delaFuenteMarcos2025Assessing, Chandler2025, Cordiner2025JWST3I, Seligman2025} and shown in Fig.~\ref{fig:jets}, the morphology of the near-nucleus coma (within a projected radius of $\sim10{,}000$~km) remained remarkably stable over several weeks, displaying a persistent, nearly sunward dust plume visible since the comet’s discovery \citep{Alarcon2025ATel17264, Jewitt2025HST3IATLAS}.
The observed fan-like structure, in the same direction of the jet we detected, can be interpreted as the projection onto the plane of the sky of an emission cone originating from a high-latitude active region on the sunlit hemisphere of the rotating nucleus.
A numerical model of the inner coma proposed by \citet{Oldani2025ATel17350} successfully reproduces the observed morphology by invoking a discrete active source located at a latitude of approximately $+75^{\circ}$, with micron-sized dust grains ejected at velocities consistent with gas drag driven by CO$_2$ sublimation.

The coma composition derived from near- and mid-infrared spectroscopy reveals a CO$_2$-dominated volatile budget 
with a CO$_2$/H$_2$O ratio significantly higher than in typical Solar System comets \citep{Cordiner2025JWST3I}. 
Such a composition promotes efficient dust entrainment and contributes to the diffuse, isotropic appearance of the coma. 
Time-series imaging further indicates that the coma morphology remained globally stable between July and September~2025, 
although subtle variations in the brightness of the sunward fan were observed, possibly modulated by the nucleus rotation 
\citep{SantanaRos2025Temporal}.

{Polarimetric measurements reveal an unusually strong negative polarization branch and a low inversion angle, indicating that the coma dust of 3I is dominated by large, compact particles. Such a dust population naturally explains the relatively weak effects of solar radiation pressure and the smooth transition between the sunward fan and the antisolar tail. As has been argued by \citet{Gray2025Polarisation3I}, the deep negative polarization branch can be reproduced by large grains composed of mixed icy and dark materials. Consistently, the pre-perihelion visible spectrum of 3I displays a moderately red slope, likely caused by scattering from large dust grains \citep{2025A&A...700L..10A}.

\subsection{Dust tail of 3I}
In addition to its nearly symmetric inner coma and prominent sunward fan, 3I exhibits a dust tail that developed progressively as the comet approached the Sun. 
By mid-August, a tail became clearly apparent in the antisolar direction \citep{Tonry2025ATLAS}. In our sample (see table \ref{table:obs}), this tail is detected in images from August~14, 18, 19, 21, 24, and~29, with measured PAs -- from isophotal contours -- of $100.9^\circ$, $102.6^\circ$, $101.3^\circ$, $100.4^\circ$, $102.6^\circ$, and $103.3^\circ$ (yellow lines in Fig.~\ref{fig:jets}, errors of $\pm 0.5^\circ$), respectively, in agreement with the antisolar direction (yellow arrows  in Fig.~\ref{fig:jets}) of $101.4^\circ$, $101.7^\circ$, $101.8^\circ$, $102.3^\circ$, and $102.7^\circ$ for the same dates. 

The feature remained short and diffuse, consistent with weak radiation-pressure effects acting on relatively large dust particles. 
This morphology suggests that the dust-size distribution is dominated by grains of several tens to hundreds of microns, which experience limited acceleration by solar photons \citep{SantanaRos2025Temporal}. 
The faint appearance and slow temporal evolution of the tail can therefore be explained by a combination of (i) low ejection velocities, (ii) large average grain size, and (iii) projection effects due to the observing geometry. 

\subsection{Rotation period of 3I}
Assuming that the comet’s aspect angle -- defined as the angle between the sky-projected spin axis and the observer’s line of sight -- remained approximately constant during the observing timespan, we can also constrain the rotation period from the temporal variability of the jet morphology. For practical purposes, this aspect angle is approximated by the angle of the target’s heliocentric velocity vector (“PsAMV” in the JPL ephemerides) as projected onto the observer’s plane of the sky. Under this approximation, the measured PAs of the two-dimensional jet projection are corrected using the corresponding PsAMV values provided by the orbit ephemerids. Over the August observing interval, the PsAMV varied by only $\sim$ $4^\circ$, indicating that the viewing geometry was relatively stable. We further assume that the nucleus rotated uniformly throughout the campaign, with no measurable precession or nutation. 

We note that the jet is not located exactly at the rotational pole. A perfectly polar jet would produce a strictly constant PA under the nearly fixed viewing geometry of our campaign. Instead, the observed PA exhibits a small but significant modulation ($\sim$ $12^\circ$ peak-to-peak; Fig. \ref{fig:phase}), indicating that the active source lies at high, but not polar, latitude compatible with the model of \citet{Oldani2025ATel17350}. This geometry naturally yields a nearly constant PA with low-amplitude oscillations, consistent with the measured values and enabling the use of PA modulation to infer the rotation period.

We searched for periodicities in the temporal evolution of the measured PAs using the phase dispersion minimization (PDM) method \citep{Stellingwerf1978PDM}.
The best-fit solution corresponds to a period of $7.74 \pm 0.35$~h (Fig.~\ref{fig:phase}), with the zero phase defined at MJD~60891.1035.
If the jet originates from a single active area located near one of the rotational poles, the true nucleus rotation period would be twice this value, i.e., $P_{\mathrm{rot}} = 15.48 \pm 0.70$~h.
This period, derived from images acquired in August, is slightly shorter than those obtained from independent photometric analyses of data taken about a month earlier,  
$P_{\mathrm{rot}} = 16.79 \pm 0.23$~h \citep{delaFuenteMarcos2025Assessing} and $P_{\mathrm{rot}} = 16.16 \pm 0.01$~h \citep{SantanaRos2025Temporal}.
The discrepancy may reflect the intrinsic uncertainties of the indirect methods used, or it may indicate an actual change in the rotation period caused by outgassing torques acting on the jet. 

From the mean value of the measured jet PAs, the sky-projected orientation of the spin axis was derived (see the horizontal red line in Fig.~\ref{fig:phase}), yielding ${\rm PA} = 280.7 \pm 0.2^{\circ}$.
Considering the full range of observed PAs, the jet defines a fan or plume extending from ${\rm PA} = 274.7 \pm 0.6^{\circ}$ to ${\rm PA} = 288.8 \pm 0.8^{\circ}$. These values are consistent with the fan orientations reported by other observers.
Early imaging obtained on July~2 with the Two-meter Twin Telescope (TTT) revealed a plume at ${\rm PA} = 279.5 \pm 1.4^{\circ}$ \citep{delaFuenteMarcos2025Assessing}, while subsequent Hubble Space Telescope (HST) observations confirmed a similar morphology, showing a well-resolved fan-like feature at ${\rm PA} = 280 \pm 10^{\circ}$ \citep{Jewitt2025HST3IATLAS}.

\begin{figure}[h!]
\centering
\includegraphics[width=\hsize]{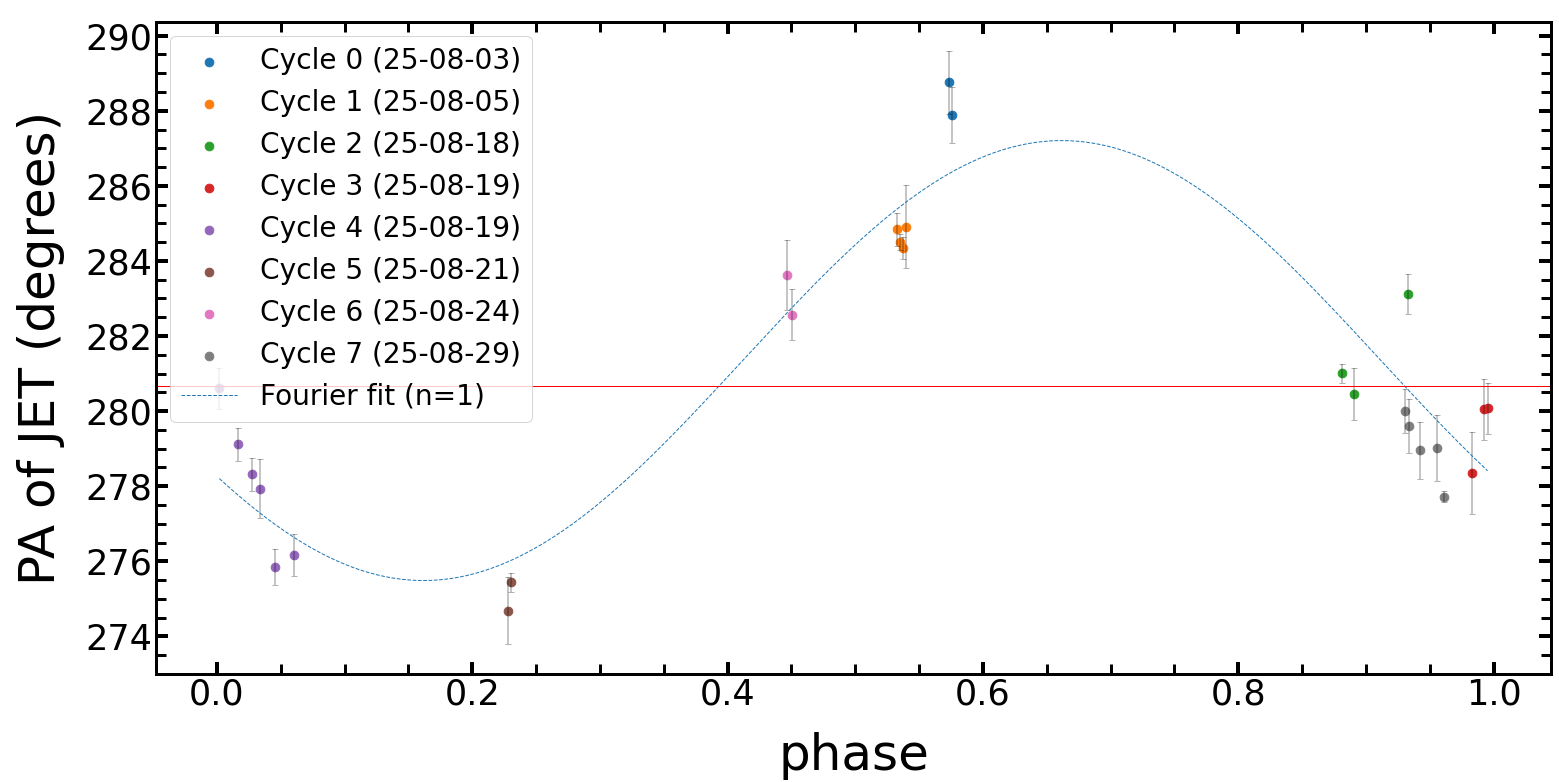}
\caption{Phase angle of the PA of jet in the inner coma of 3I, measured at 6000~km from the cometary optocenter and phased with the calculated $7.74 \pm 0.35$~h period. The horizontal red line is the mean value of PA angles representing the sky-projected orientation of the spin axis.}
\label{fig:phase}
\end{figure}

\section{Conclusions} \label{sec:conclu}

In this Letter, we have presented an extensive monitoring campaign of 3I conducted over 37 nights between 2025,~July~2 and~September~5 with the imaging capabilities of the Two-meter Twin Telescope (TTT), aimed at characterizing the morphology of its coma.  
Our main conclusions are as follows:

\begin{enumerate}

\item The morphology of the coma evolves from a sunward dust fan in pre-August observations to a more pronounced antisolar tail at later dates (Fig. \ref{fig:jets}). This delayed development of the tail likely results from the weak influence of solar radiation pressure on the slow-moving coma dust.

\item A Laplacian-filtering technique \citep{SerraRicart2015Rotation} applied to the combined images revealed fine-scale structures in the inner coma. 
A faint but well-defined jet was detected on seven nights (2025, August~3, 5, 18, 19, 21, 24, and~29).
Position angle measurements at a projected distance of 6000~km from the cometary optocenter exhibit a periodic modulation centered at $\sim280\degr$ (Fig. \ref{fig:phase}), consistent with a high-latitude jet undergoing precessional motion around the sky-projected spin axis. This is the first periodic jet-angle modulation detected in an interstellar comet. These jet-angle oscillations therefore offer a purely morphological route to constraining the spin axis. From the mean value of jet PAs, the sky-projected orientation of the spin axis was determined to be $\mathrm{PA} = 280.7 \pm 0.2^{\circ}$.

\item 
The derived periodicity of the jet PA of  $7.74 \pm 0.35$~h may imply a nucleus rotation period of $P_{\mathrm{rot}} = 15.48 \pm 0.70$~h, slightly shorter than the value of $16.79 \pm 0.23$~h reported by \citet{delaFuenteMarcos2025Assessing} from the inner-coma photometry.

\end{enumerate}

\begin{acknowledgements}
This article is based on observations made in the 
Two-meter Twin Telescope (TTT\footnote{\url{http://ttt.iac.es}}) sited at the Teide Observatory of the Instituto 
de Astrofísica de Canarias (IAC), that Light Bridges operates in Tenerife, Canary Islands (Spain). The observation time rights (DTO) used for this research were consumed in the PEI "PLANETIX25". This research used storage and 
computing capacity in ASTRO POC's EDGE computing center at Tenerife under the form of Indefeasible Computer Rights 
(ICR). The ICR were consumed in the PEI "PLANETIX25" with the collaboration of Betchle AG. 
Dr. Antonio Maudes’s insights in economics and law were instrumental in shaping the development of this work.
JL acknowledge support from the Agencia Estatal de Investigaci\'on del Ministerio de Ciencia e Innovaci\'on (AEI-MCINN) under grant "Hydrated Minerals and Organic Compounds in Primitive Asteroids" with reference PID2020-120464GB-100. 
\end{acknowledgements}

\bibliographystyle{aa}
\bibliography{3IATLAS}

\begin{appendix}

\begin{table*}[h!]
\section{Observations and Laplacian-filtering process}
3I was extensively monitoring on 37 nights between 2025 July 2 and 2025 September 9 using the imaging capabilities of TTT3 facility (see section \ref{sec:observations} for details). Observational circumstances are summarized in Table \ref{table:obs}. Figures \ref{fig:filunfil} and \ref{fig:fitting} show the Laplacian-filtering technique and the method to calculate the PA of the jet, respectively.

\caption{Observing log for 3I/ATLAS}                 
\label{table:obs}    
\centering                        
\begin{tabular}{c c c c c c c c c}      
\hline\hline               
Date & UTC$_{\rm start}$ & UTC$_{\rm end}$ & Exp. time (s) & N. images & Seeing ($''$) & Airmass & $r$ (au) \\         
\hline                      
              2025-07-02 & 2025-07-02 21:28 & 2025-07-03 02:47 &                11950 &           239 &             1.1 &         1.5–2.0 &       4.48 \\
              2025-07-03 & 2025-07-03 23:46 & 2025-07-04 03:42 &                10440 &           232 &             1.1 &         1.5–2.0 &       4.41 \\
              2025-07-05 & 2025-07-06 03:22 & 2025-07-06 04:04 &                  500 &            10 &             1.9 &         1.6–1.7 &       4.35 \\
              2025-07-06 & 2025-07-06 23:56 & 2025-07-07 04:22 &                 1400 &            28 &             1.9 &         1.5–1.9 &       4.32 \\
              2025-07-07 & 2025-07-07 23:31 & 2025-07-07 23:46 &                  750 &            15 &             1.6 &         1.8–1.9 &       4.28 \\
              2025-07-11 & 2025-07-11 23:41 & 2025-07-12 04:34 &                10350 &           207 &             1.3 &         1.5–2.3 &       4.15 \\
              2025-07-12 & 2025-07-12 23:44 & 2025-07-13 04:30 &                 5900 &           118 &             1.5 &         1.5–2.4 &       4.11 \\
              2025-07-13 & 2025-07-14 00:28 & 2025-07-14 04:05 &                 7600 &           152 &             1.6 &         1.5–2.1 &       4.08 \\
              2025-07-14 & 2025-07-15 00:13 & 2025-07-15 04:16 &                 2150 &            43 &             1.6 &         1.5–2.3 &       4.05 \\
              2025-07-17 & 2025-07-18 01:03 & 2025-07-18 01:05 &                  150 &             3 &             1.9 &         1.5–1.5 &       3.95 \\
              2025-07-18 & 2025-07-18 23:32 & 2025-07-19 01:10 &                 1850 &            37 &             1.7 &         1.5–1.6 &       3.92 \\
              2025-07-19 & 2025-07-19 23:32 & 2025-07-20 02:03 &                 1750 &            35 &             1.6 &         1.5–1.6 &       3.89 \\
              2025-07-20 & 2025-07-20 23:29 & 2025-07-21 02:01 &                 4750 &            95 &             1.4 &         1.5–1.6 &       3.85 \\
              2025-07-21 & 2025-07-21 23:28 & 2025-07-22 02:00 &                 2700 &            54 &             1.2 &         1.5–1.6 &       3.82 \\
              2025-07-24 & 2025-07-25 01:33 & 2025-07-25 01:33 &                   30 &             1 &             1.6 &         1.6–1.6 &       3.72 \\
              2025-07-25 & 2025-07-25 23:30 & 2025-07-25 23:34 &                   60 &             2 &             1.4 &         1.5–1.5 &       3.69 \\
              2025-07-26 & 2025-07-26 23:25 & 2025-07-27 01:29 &                 2950 &            59 &             1.7 &         1.4–1.6 &       3.66 \\
              2025-07-30 & 2025-07-30 23:23 & 2025-07-31 02:00 &                 3400 &            68 &             1.9 &         1.4–2.0 &       3.53 \\
              2025-08-02 & 2025-08-02 23:18 & 2025-08-03 02:16 &                 7950 &           159 &             1.3 &         1.4–2.4 &       3.43 \\
              2025-08-03 & 2025-08-03 23:06 & 2025-08-04 02:09 &                 8200 &           164 &             1.5 &         1.4–2.3 &       3.40 \\
              2025-08-04 & 2025-08-04 23:04 & 2025-08-05 01:55 &                 6050 &           121 &             1.7 &         1.4–2.2 &       3.36 \\
              2025-08-05 & 2025-08-05 23:04 & 2025-08-06 01:57 &                 7500 &           150 &             1.6 &         1.4–2.4 &       3.33 \\
              2025-08-06 & 2025-08-06 23:14 & 2025-08-07 01:51 &                 4360 &            88 &             1.4 &         1.4–2.4 &       3.30 \\
              2025-08-18 & 2025-08-18 23:01 & 2025-08-18 23:59 &                 2460 &            50 &             1.1 &         1.6–1.9 &       2.92 \\
              2025-08-19 & 2025-08-19 23:02 & 2025-08-20 00:29 &                 2960 &            60 &             1.3 &         1.6–2.3 &       2.89 \\
              2025-08-20 & 2025-08-20 23:01 & 2025-08-20 23:53 &                 2160 &            44 &             1.6 &         1.6–2.0 &       2.86 \\
              2025-08-21 & 2025-08-21 22:58 & 2025-08-21 23:55 &                 2360 &            48 &             1.5 &         1.6–2.1 &       2.82 \\
              2025-08-22 & 2025-08-22 22:54 & 2025-08-22 23:41 &                 1860 &            38 &             1.7 &         1.6–2.0 &       2.79 \\
              2025-08-24 & 2025-08-24 22:41 & 2025-08-24 23:59 &                 2110 &            43 &             1.4 &         1.6–2.4 &       2.73 \\
              2025-08-25 & 2025-08-25 22:48 & 2025-08-25 22:50 &                  100 &             2 &             1.9 &         1.7–1.7 &       2.70 \\
              2025-08-26 & 2025-08-26 22:59 & 2025-08-26 23:15 &                  200 &             4 &             2.0 &         1.8–2.0 &       2.67 \\
              2025-08-27 & 2025-08-27 22:46 & 2025-08-27 23:40 &                 2100 &            42 &             1.7 &         1.8–2.3 &       2.64 \\
              2025-08-28 & 2025-08-28 22:37 & 2025-08-28 23:34 &                 1960 &            40 &             1.3 &         1.8–2.3 &       2.61 \\
              2025-08-29 & 2025-08-29 22:39 & 2025-08-29 23:21 &                 1360 &            28 &             1.4 &         1.8–2.2 &       2.58 \\
              2025-09-01 & 2025-09-01 23:01 & 2025-09-01 23:05 &                   60 &             2 &             1.5 &         2.2–2.3 &       2.49 \\
              2025-09-02 & 2025-09-02 22:34 & 2025-09-02 22:38 &                   60 &             2 &             1.4 &         2.0–2.0 &       2.46 \\
              2025-09-05 & 2025-09-05 22:43 & 2025-09-05 22:43 &                   30 &             1 &             1.8 &         2.3–2.3 &       2.37 \\
\hline                                  
\end{tabular}
\tablefoot{The table includes the date of the observations, the starting and end times (UTC$_{\rm start}$ and UTC$_{\rm end}$), the total exposition time in seconds, the total number of images, median seeing, range of airmass and cometary distance to the Sun ($r$).}
\end{table*}

\begin{figure*}[h!]
\centering
\resizebox{15cm}{8.87cm}
{\includegraphics {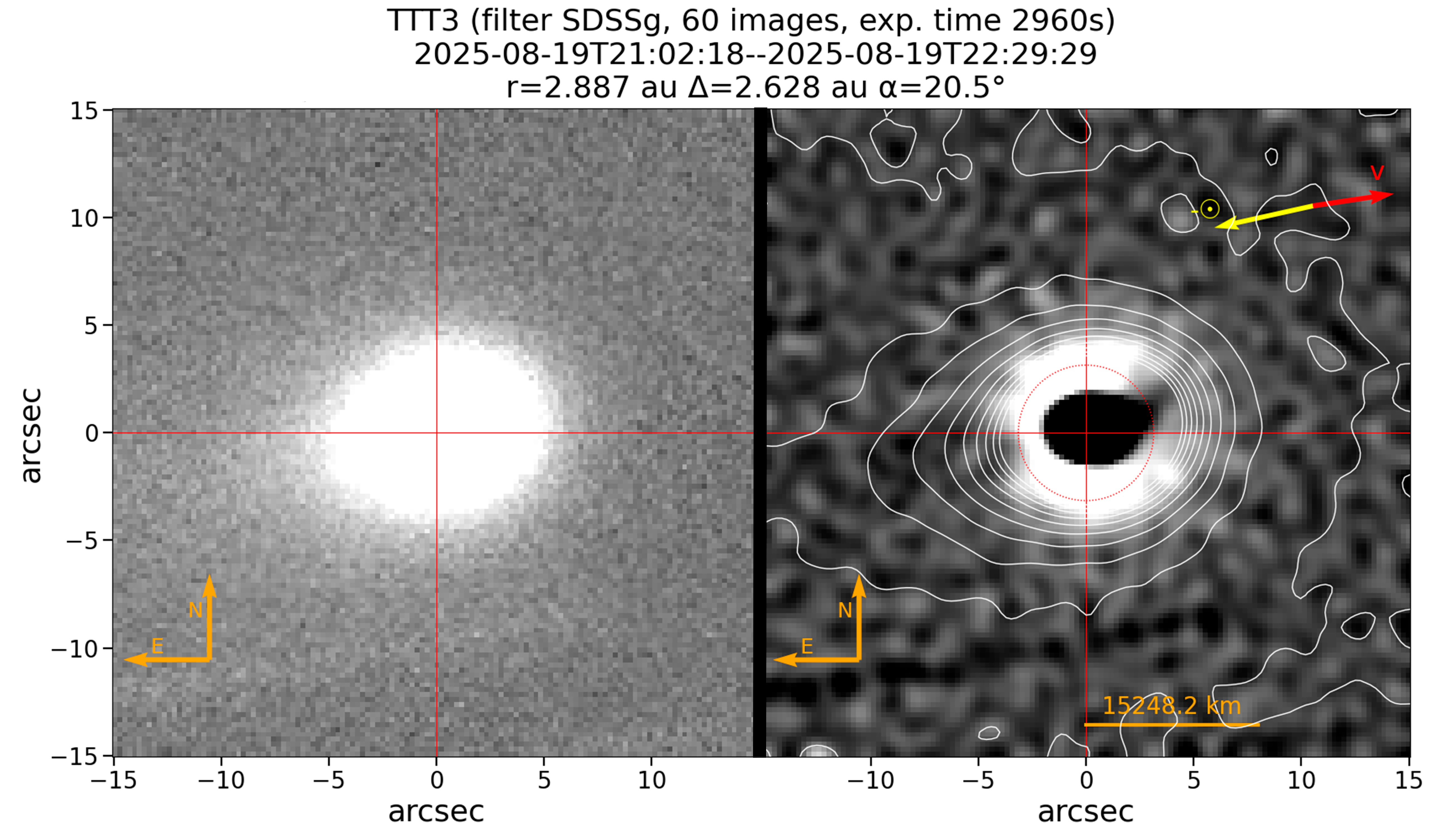}}
\caption{
Unprocessed image (left) and Laplacian-filtered image (right) of the inner coma of the interstellar comet 3I/ATLAS (C/2025 N1). The observation date (YYYY–MM–DD), start and end times (UTC), number of sidereal-tracking exposures, and total integration time are indicated above the panels. The projected velocity vector (red arrow) and antisolar direction (yellow arrow) are shown, together with the image scale and orientation. Red crosshairs mark the comet’s optocenter.
Isophotal contours of the original (unfiltered) frame are overplotted on the filtered image (right) using ten logarithmically spaced levels between the 20th and 95th percentiles of the pixel-intensity distribution. The jet refers to the narrow, linear, black feature extending roughly northwest in the filtered image. It originates on the sunward side of the nucleus, in contrast to the dust tail, which points in the antisolar direction.
The red dotted circle marks a projected radial distance of 6000 km from the optocenter, at which the jet PA is measured.}

\label{fig:filunfil}
\end{figure*}

\begin{figure*}[h!]
\centering
\includegraphics[width=\hsize]{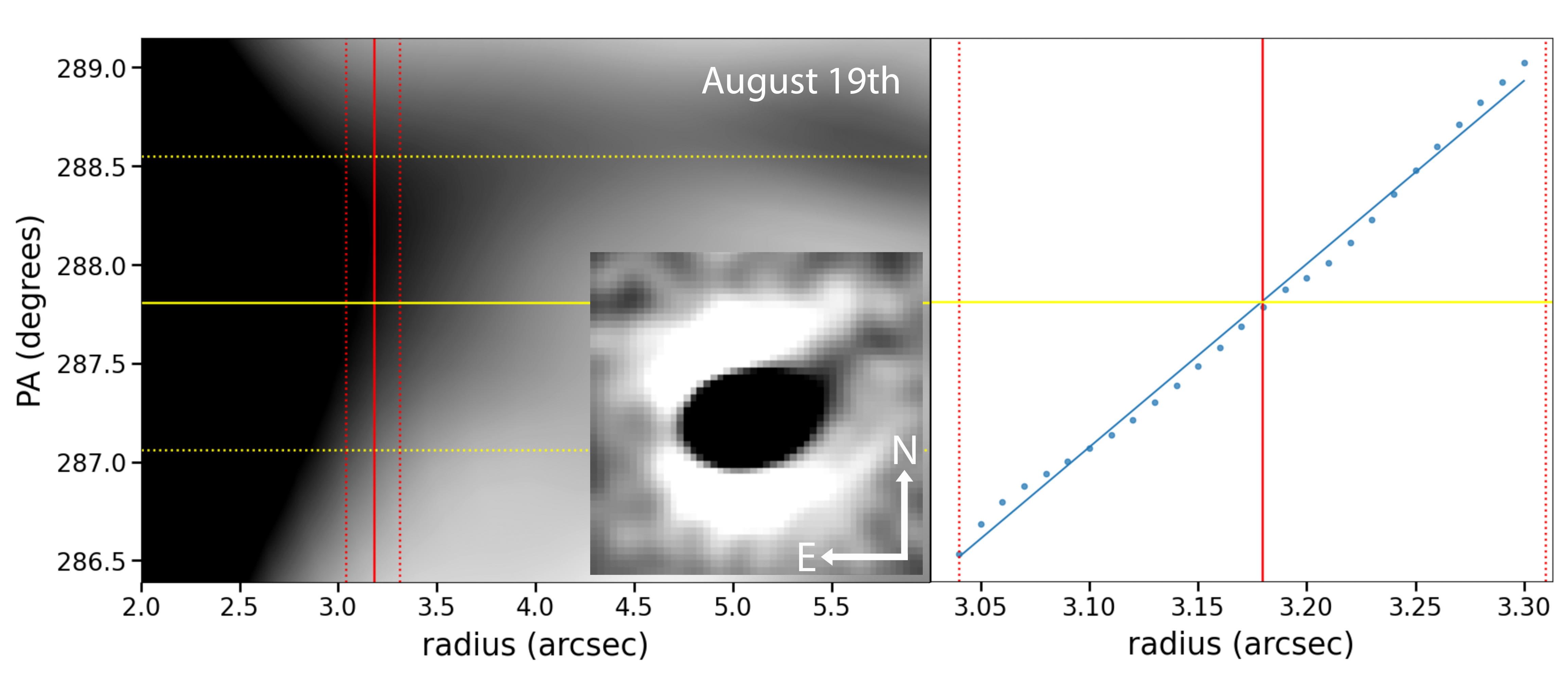}
\caption{Left: Polar Laplacian-filtered image of 3I from the August~19 observations (the inset shows the corresponding sky-plane filtered image).
From the polar-filtered frames, PA profiles were extracted at several radial distances within $\pm250$~km (dotted red lines) around 6000~km from the optocenter (solid red line).
At each radius, the PA of the jet was measured by fitting a parabola to the local intensity minimum in the jet region.
Right: Linear regression of the measured PAs as a function of radius yields the final PA jet value (yellow line) with errors (dotted yellow lines) at $r = 6000$~km.}
\label{fig:fitting}
\end{figure*}

\end{appendix}

\end{document}